\documentclass[a4paper,11pt]{article}

 \usepackage{amsfonts,amsthm,color,amssymb}
 \usepackage{latexsym,epsf}

\newcommand{\eq}{\begin{equation}}
\newcommand{\en}{\end{equation}}
\newcommand{\eqa}{\begin{eqnarray}}
\newcommand{\ena}{\end{eqnarray}}

\newcommand{\ha}{\hat{a}}
\newcommand{\hb}{\hat{b}}
\newcommand{\hc}{\hat{c}}
\newcommand{\hd}{\hat{d}}

\newcommand{\wb}{\widetilde{B}}
\newcommand{\wm}{\widetilde{M}}
\newcommand{\wt}{\widetilde{T}}
\newcommand{\pw}{\widetilde{P}}

\begin{document}

\setlength{\unitlength}{1mm} \thispagestyle{empty}

\vspace*{0.1cm}


\begin{center}
{\small \bf GHZ States, Almost-Complex Structure and Yang--Baxter Equation (I) \\[2mm]}

\vspace{.5cm}

 Yong Zhang${}^{ab}$\footnote{yong@physics.utah.edu}
 and Mo-Lin Ge${}^b$\footnote{geml@nankai.edu.cn}\\[.1cm]

 ${}^a$ Department of Physics, University of Utah \\
  115 S, 1400 E, Room 201, Salt Lake City, UT 84112-0830
\\[0.1cm]

 ${}^b$ Theoretical Physics Division, Chern Institute of Mathematics \\
  Nankai University, Tianjin 300071, P. R. China\\[0.1cm]

\end{center}

\vspace{0.2cm}

\begin{center}
\parbox{12cm}{
\centerline{\small  \bf Abstract} \small \noindent

Recent study suggests that there are natural connections between
quantum information theory and the Yang--Baxter equation. In this
paper, in terms of the generalized almost-complex structure and with
the help of its algebra, we define the generalized Bell matrix to
yield all the GHZ states from the product base, prove it to form a
unitary braid representation and present a new type of solution of
the quantum Yang--Baxter equation. We also study Yang-Baxterization,
Hamiltonian, projectors, diagonalization, noncommutative geometry,
quantum algebra and FRT dual algebra associated with this
generalized Bell matrix.

 }

\end{center}

\vspace*{10mm}
\begin{tabbing}

\end{tabbing}

\newpage

\section{Introduction}

Recently, a series of papers
 \cite{dye,kauffman1,yong1,yong2,wang,yong3,yong4,yong5,yong6,yong7}
 have suggested there are natural and deep connections between quantum
 information theory \cite{nielsen} and the Yang--Baxter equation (YBE)
\cite{yang,baxter}. Unitary solutions of the braided YBE (i.e., the
braid group relation) \cite{dye,kauffman1} as well as unitary
solutions of the quantum Yang--Baxter equation (QYBE)
\cite{yong1,yong2} can be often identified with universal quantum
gates \cite{bb}. Yang--Baxterization \cite{jones} is exploited to
set up the Schr{\"o}dinger equation determining the unitary
evolution of a unitary braid gate \cite{yong1,yong2}. Furthermore,
the Werner state \cite{werner} is viewed as a rational solution of
the QYBE and the isotropic state \cite{horodecki} with a specific
parameter forms a braid representation, see \cite{yong4, yong5}.
More interestingly, the Temperley--Lieb algebra \cite{lieb} deriving
a braid representation in the state model for knot theory
\cite{kauffman2} is found to present a suitable mathematical
framework for a unified description of various kinds of quantum
teleportation phenomena \cite{bennett}, see \cite{yong6,yong7}.

The present paper is a further extension of the previous published
research work \cite{yong1,yong2,yong3} in which the Bell matrix has
been recognized to form a unitary braid representation and generate
all the Bell states from the product base. In this paper, a unitary
braid representation also called the Bell matrix for convenience is
defined to create all the Greenberger-Horne-Zeilinger states (GHZ
states) from the product base. The GHZ states are maximally
multipartite entangled states (a natural generalization of the Bell
states) and play important roles in the study of quantum information
phenomena \cite{ghz1,ghz2,ghz3}. More importantly, this Bell matrix
has a form in terms of the almost-complex structure which is
fundamental for complex and K{\"a}hler geometry and symplectic
geometry. Therefore, our paper is building heuristic connections
among quantum information theory, the Yang--Baxter equation and
differential geometry.

We hereby summarize our main result which is new to our knowledge.

 \begin{enumerate}

\item We define the Bell matrix to produce all the GHZ
states from the product base, prove it to be a unitary braid
representation, and derive the Hamiltonian to determine the
 unitary evolution of the GHZ states.

\item We recognize the almost-complex structure in the formulation
of the Bell matrix as well as its algebra in the proof for the Bell
matrix satisfying the braided YBE, and exploit it to represent a new
type of the solution of the QYBE.

 \item We study topics associated with the generalized Bell matrix which
 include  Yang--Baxterization, diagonalization, noncommutative
 geometry, quantum algebra via the $RTT$ relation and standard FRT
 procedure \cite{rtf,faddeev}.

\end{enumerate}

As the first paper in this research project, for simplicity, the
present manuscript only focuses on the generalized Bell matrix of
the type $2^{2n}\times 2^{2n}$ corresponding to the GHZ states of an
even number of objects, while our result on the generalized Bell
matrix of the type $2^{2n+1}\times 2^{2n+1}$ is collected
\cite{yong8}.

The plan of this paper is organized as follows. Section 2 sketches
the definition of the GHZ states and represent the Bell matrix in
terms of the almost complex structure. Section 3 introduces the
generalized Bell matrix and show both an algebra and a interesting
type of solution of the QYBE in terms of the generalized
almost-complex structure. Sections 4 and 5 briefly deal with various
 topics about the generalized Bell matrix: projectors, diagonalization,
 noncommutative geometry, quantum algebra and FRT dual algebra. Last
 section concludes with worthwhile problems for further
research.

\section{GHZ states, Bell matrix and Hamiltonian}

This section is devised to set up a simplest example to be
appreciated by readers mostly interested in quantum information and
physics, and it explains how to observe the Bell matrix from the
formulation of the GHZ states (as well as the almost-complex
structure from the Bell matrix) and how to obtain Hamiltonians to
determine the unitary evolution of the GHZ states.

\subsection{GHZ states, Bell matrix and almost-complex structure}

In the $2^N$-dimensional Hilbert space with the base denoted by the
Dirac kets $|m_1,m_2,\cdots,m_N\rangle$, $m_1,\cdots,m_N=\pm \frac 1
2$, there are $2^N$ linearly independent GHZ states of $N$-objects
having the form \eq
 \frac 1 {\sqrt 2} (|m_1,m_2,\cdots,m_N\rangle \pm |-m_1,-m_2,\cdots,-m_N\rangle
 )\en which are maximally entangled states in quantum information
 theory \cite{nielsen}. In this paper, all the GHZ states are found to
 be generated by the Bell matrix acting on the chosen product base,
 \eq
 |\Phi_k\rangle=|m_1,m_2,\cdots,m_N\rangle, \,\,\,\, |\Phi_{2^N-k+1}\rangle
  =|-m_1,-m_2,\cdots,-m_N\rangle,
 \en
where $1\le k \le 2^{N-1}$. One can take a notation similar to
\cite{fujii1,fujii2}, \eq k[m_1,\cdots,m_N]=2^{N-1}+\frac 1
2-\sum_{i=1}^N 2^{N-i}\,\, m_i \en which has the result at $N=2$,
for example, \eq k[\frac 1 2,\frac 1 2]=1,\,\, k[\frac 1 2,-\frac 1
2]=2,\,\, k[-\frac 1 2,\frac 1 2]=3,\,\,k[-\frac 1 2,-\frac 1 2]=4,
\en assigned to label the GHZ states of two objects (the well known
Bell states).

The $4\times 4$ Bell matrix $B_4$
 acts on the product base
 $|\frac 1 2\frac 1 2\rangle$,
$|\frac {1} 2 \frac {-1} 2\rangle$ and $|\frac {-1} 2 \frac 1
2\rangle$, $|\frac {-1} 2\frac {-1} 2\rangle$ to yield the Bell
states, and it has a known form
\cite{dye,kauffman1,yong1,yong2,yong3}, \eq
 B_{4}=(B_{kn,\,lm})_4 =\frac 1 {\sqrt 2}\left(
 \begin{array}{cccc}
 1 & 0 & 0 & 1 \\
 0 & 1 & 1 & 0 \\
 0 & -1 & 1 & 0 \\
 -1 & 0 & 0 & 1 \\
 \end{array} \right),\qquad k,n,l,m=\frac 1 2, -\frac 1 2,
\en and the $8\times 8$ Bell matrix $B_8$ given by \eqa &&
B_{8}\equiv (B_{\alpha l,\,\beta m})_8 = \frac 1 {\sqrt
2}\left(\begin{array}{llllllll}
 1 & 0 & 0 & 0 & 0 & 0 & 0 & 1 \\
  0 & 1 & 0 & 0 & 0 & 0 & 1 & 0 \\
  0 & 0 & 1 & 0 & 0 & 1 & 0 & 0 \\
   0 & 0 & 0 & 1 & 1 & 0 & 0 & 0 \\
  0 & 0 & 0 & -1 & 1 & 0 & 0 & 0 \\
  0 & 0 & -1 & 0 & 0 & 1 & 0 & 0 \\
  0 & -1 & 0 & 0 & 0 & 0 & 1 & 0 \\
   -1 & 0 & 0 & 0 & 0 & 0 & 0 & 1 \\
\end{array} \right),
 \nonumber\\
&& \alpha, \beta=\frac 3 2,\frac 1 2, -\frac 1 2, -\frac 3 2,\,\,
l,m=\frac 1 2, -\frac 1 2 \ena creates the GHZ states of three
objects by acting on $|\Phi_k\rangle$, $1\le k \le 8$.

The $2^N\times 2^N$ Bell matrix generating the GHZ states of
$N$-objects from the product base $|\Phi_k\rangle$, $1\le k \le
2^N$, has a form in terms of the almost-complex structure
\footnote{The almost-complex structure is usually denoted by the
symbol $J$ in the literature and it is a linear map from a real
vector space to itself satisfying $J^2=-1$. More details on geometry
underlying what we are presenting here will be discussed elsewhere.}
denoted by $M$, \eq
 B=\frac 1 {\sqrt 2} (1\!\! 1 +M),  \qquad B_{ij,kl}\equiv
 B_{ij}^{kl}=\frac 1 {\sqrt 2} (\delta_i^k\delta_j^l +M_{ij}^{kl}) \en
 where $1\!\! 1$ denotes the identity matrix, the lower index of
 $B_{2^N}$ is omitted for convenience, $\delta_i^j$ is the Kronecker
 function of two variables $i,j$, which is $1$ if $i=j$ and $0$
 otherwise, and the almost-complex structure $M$ has the component
 formalism using the step function $\epsilon(i)$,
 \eq
 M_{ij,kl} \equiv M_{ij}^{kl}= \epsilon(i)\delta_i^{-k}\delta_j^{-l}, \qquad
 \epsilon(i)=1, i\ge0;\,\,\, \epsilon(i)=-1, i<0,
 \en
which satisfies $M^2=-1\!\!1$. In terms of the tensor product of the
Pauli matrices, the Bell matrix $B$ and the almost complex structure
$M$ for $N$-objects have the forms given by
 \eq  B=e^{\frac \pi 4 M },\,\,
 M= \sqrt{-1} \sigma_y\otimes (\sigma_x)^{\otimes (N-1)},\,\,
 (\sigma_x)^{\otimes (N-1)} =\underbrace{\sigma_x\otimes \cdots \otimes
 \sigma_x}_{N-1}.
 \en

 Note that there exist other interesting matrices related to the GHZ
 states, for example, one can have matrix entries $\epsilon(i) B_{ij,kl}$
 for a new matrix. But so far as the authors know, only the Bell matrix
 is found to form a unitary braid representation.

\subsection{Yang--Baxterization and Hamiltonian}

The Bell matrix $B$ satisfies the following characteristic equation
given by
 \eq
 \label{char}
(B-\frac {1+\sqrt{-1}} {\sqrt 2}1\!\! 1)(B-\frac {1-\sqrt{-1}}
{\sqrt 2}1\!\! 1)=0
 \en
which suggests it having two distinct eigenvalues $\frac {1\pm
\sqrt{-1}} {\sqrt 2} $. Using Yang--Baxterization\footnote{See
\cite{yong2} or Subsection 3.1 and Subsection 4.1 for the detail.
Here Yang--Baxterization is applied to the Bell matrix of the type
$2^{2n}\times 2^{2n}$, while Yang--Baxterization of the Bell matrix
of the type $2^{2n+1}\times 2^{2n+1}$ is rather subtle to be
presented \cite{yong8}.}, a solution of the QYBE with the Bell
matrix as its asymptotic limit, is obtained to be
 \eq
 \label{rx}
 \check{R}(x)=B+x B^{-1}=\frac 1 {\sqrt 2} (1+x)1\!\!
 1+\frac 1 {\sqrt 2}(1-x)M.
 \en
As this solution $\check{R}(x)$ is required to be unitary, it needs
a normalization factor $\rho$ with a real spectral parameter $x$,
 \eq
B(x)=\rho^{-\frac 1 2}\check{R}(x), \qquad \rho= 1+x^2,\,\,
  x\in {\mathbb R}.
 \en

As the real spectral parameter $x$  plays the role of the time
variable, the Schr{\"o}dinger equation describing the unitary
evolution of a state $\phi$ (independent of $x$) determined by the
$B(x)$ matrix, i.e., $\psi(x)=B(x)\phi$, has the form
 \eq
 \label{scr1}
 \sqrt{-1}\frac {\partial} {\partial x}\psi(x)=H(x)\psi(x),\qquad
  H(x)\equiv \sqrt{-1}\frac {\partial B(x)} {\partial x}  B^{-1}(x),
 \en
where the time-dependent Hamiltonian $H(x)$ is  given by
 \eq
H(x)= \sqrt{-1} \frac {\partial } {\partial x}(\rho^{-\frac 1 2}
{\check R}(x))(\rho^{-\frac 1 2}
 \check{R}(x))^{-1}=- \sqrt{-1}  \rho^{-1} M.
\en To construct the time-independent Hamiltonian, a new time
variable $\theta$ instead of the spectral parameter $x$ is
introduced in the way \eq \cos\theta=\frac 1 {\sqrt{1+x^2}}, \qquad
 \sin\theta=\frac x {\sqrt{1+x^2}},
 \en
so that the Bell matrix $B(x)$ has a new formulation as a function
of $\theta$,
 \eq
 B(\theta)=\cos\theta B+\sin\theta B^{-1}=e^{(\frac \pi 4-\theta)
 M},
  \en
and hence the Schr{\"o}dinger equation for the time evolution of
$\psi(\theta)=B(\theta)\phi$ has the form
 \eq
 \label{scr2}
 \sqrt{-1}\frac {\partial} {\partial
 \theta}\psi(\theta)=H\psi(\theta),\qquad
 H\equiv\sqrt{-1}\frac {\partial B(\theta)}
 {\partial \theta}  B^{-1}(\theta)=-\sqrt{-1} M,
 \en
where the time-independent Hamiltonian \footnote{The Hamiltonian
used in our previous published papers \cite{yong1,yong2} has an
additional numerical factor $\frac 1 2$ compared to the
time-independent Hamiltonian (\ref{scr2}). This factor $\frac 1 2$
is very important when we recognize the action of the four
dimensional unitary evolution operator $\exp{\frac 1 2 \theta M}$ on
the product base to be equivalent to a product of two unitary
rotations of Wigner functions for the Bell states $|\frac 1 2 \frac
1 2\rangle \pm |\frac {-1} 2 \frac {-1} 2 \rangle$ and $|\frac 1 2
\frac {-1} 2\rangle \pm |\frac {-1} 2 \frac 1 2\rangle$,
respectively. Note that no boundary conditions have been imposed on
the Schr{\"o}dinger equations (\ref{scr1}) and (\ref{scr2}). } $H$
is hermitian due to the anti-hermitian of the almost-complex
structure, i.e., $M^\dag=-M$, and the unitary evolution operator
$U(\theta)$ has the form $U(\theta)=e^{-M\theta}$.

\section{Generalized Bell matrix and YBE}

This section proves the generalized\footnote{Here ``generalized"
means that the object has deformation parameters.} Bell matrix $\wb$
of the type $2^{2n}\times 2^{2n}$ to form a unitary braid
representation with the help of the algebra generated by the
generalized almost-complex structure $\wm$, and presents an
interesting type of solution of the QYBE in terms of $\wm$ which may
be not well noticed before in the literature.

\subsection{YBE and Yang--Baxterization}

In this paper, the braid group representation $\sigma$-matrix and
the QYBE solution $\check{R}(x)$-matrix are $d^2 \times d^2$
matrices acting on $ V\otimes V$ where $V$ is a $d$-dimensional
complex vector space. As $\sigma$ and $\check{R}$ act on the tensor
product $V_i\otimes V_{i+1}$, they are denoted by $\sigma_i$ and
$\check{R}_i$, respectively.

The generators $\sigma_i$ of the braid group $B_n$ satisfy the
algebraic relation called the braid group relation,
 \eqa
 \label{bgr}
   \sigma_{i}  \sigma_{i+1} \sigma_{i} &=& \sigma_{i+1} \sigma_{i} \sigma_{i+1},
   \qquad 1 \leq i \leq n-1, \nonumber\\
   \sigma_{i} \sigma_{j} &=& \sigma_{j} \sigma_{i}, \qquad
   | i - j | > 1.
\ena while the quantum Yang--Baxter equation (QYBE) has the form \eq
\label{qybe}
 \check{R}_i(x)\,\check{R}_{i+1}(x y)\,\check{R}_i(y)=
 \check{R}_{i+1}(y)\,\check{R}_i(x y)\,\check{R}_{i+1}(x)\en with
 the spectral parameters $x$ and $y$. In addition,
 the component formalism the QYBE (or the braid group relation) can be
 shown in terms of matrix entries,
 \eq
 {\check R}(x)_{i_1 j_1}^{i^\prime j^\prime}
 {\check R}(x y)_{j^\prime k_1}^{k^\prime k_2}
 {\check R}(y)_{i^\prime k^\prime}^{i_2 j_2} =
 {\check R}(y)_{j_1 k_1}^{j^\prime k^\prime}
 {\check R}(x y)_{i_1 j^\prime}^{i_2 i^\prime}
 {\check R}(x)_{i^\prime k^\prime}^{j_2 k_2}.
  \en

 In view of the fact that $\check{R}(x=0)$ forms a braid representation,
 the braid group relation is also called the braided YBE. Concerning relations
 between braid representations and
 $x$-dependent solutions of the QYBE (\ref{qybe}),
 the procedure of constructing the $\check{R}(x)$-matrix from a given braid
 representation $\sigma$-matrix is called Baxterization  \cite{jones} or
 Yang--Baxterization. For a braid representation $\sigma$ with two
 distinct eigenvalues $\lambda_1$ and $\lambda_2$, the corresponding
 $\check{R}(x)$-matrix obtained via Yang--Baxterization has the form
 \eq
 \check{R}(x)=\sigma +x \lambda_1\lambda_2 \sigma^{-1}
 \en
which has been exploited in Subsection 2.2, see (\ref{rx}).

\subsection{Unitary generalized Bell matrix as a solution of YBE}

The generalized Bell matrix $\wb$ has the form in terms of the
generalized almost-complex structure $\wm$ with deformation
parameters $q_{ij}$,
 \eq
 \label{bellcomp}
  \widetilde{B}_{ij}^{kl}=\frac 1 {\sqrt 2} (\delta_i^k\delta_j^l +
  \widetilde{M}_{ij}^{kl}), \qquad
  \widetilde{M}_{ij}^{kl}=\epsilon(i)q_{ij}\delta_i^{-k}\delta_j^{-l},
 \en
where $q_{ij}q_{-i-j}=1$ is required for $\widetilde{M}^2=-1\!\! 1$
and the step function $\epsilon(i)$ has the properties given by \eq
\epsilon(i)\epsilon(i)=1,\,\, \epsilon(i)\epsilon(-i)=-1,\,\,
\epsilon(i)=\pm 1, \en

Let $\wb$ be labeled by familiar indices by the angular momentum
theory in quantum mechanics, \eq
 (\widetilde{B}^{J_1J_2})^{b\nu}_{\mu a},\,\,
 \mu,\nu=J_1,J_1-1,\cdots,-J_1,\,\,
 a, b=J_2,J_2-1, \cdots, -J_2,
\en where $\wb^{JJ}$ denotes the generalized Bell matrix $\wb$
associated with the GHZ states of an even number of objects, for
example, \eq  \label{nj}  \widetilde{B}_{4}=\widetilde{B}^{\frac 1 2
\frac 1 2},\,\, \widetilde{B}_{16}=\widetilde{B}^{\frac 3 2 \frac 3
2},\,\, \widetilde{B}_{64}=\widetilde{B}^{\frac 7 2 \frac 7 2},
 \en
but the same type of generalized Bell matrix may be labeled
differently, for example, both $\wb^{\frac 1 2\frac 3 2}$ and
$\wb^{\frac 3 2\frac 1 2}$ are the type of $\wb_8$.

In the following, we focus on the generalized Bell matrix of the
type $\wb^{JJ}$ denoted by $\wb$, while we submit our result on the
generalized Bell matrix of the type $\wb^{J_1J_2}$, $J_1\neq J_2$ to
\cite{yong8}.

In the proof for $\wb^{JJ}$ forming a braid representation
(\ref{bgr}) in terms of its component formalism (\ref{bellcomp}),
deformation parameters $q_{ij}$ are found to
 satisfy equations,
 \eqa
 \label{qeq1}
 & & q_{i_1 j_1} q_{-i_1 -j_1} = q_{j_1 k_1} q_{-j_1 -k_1},\qquad
  i_1, j_1, k_1=J, J-1,\cdots, -J, \nonumber\\
 & & q_{j_1 k_1}=q_{i_1 j_1} q_{-j_1 k_1}q_{-i_1 j_1}, \qquad
 q_{i_1 j_1} = q_{j_1 k_1}q_{i_1 -j_1}q_{j_1 -k_1},
 \ena
 where no summation is imposed between same lower indices and
 which can be simplified by $q_{i_1 j_1} q_{-i_1 -j_1} =1$.
Furthermore, the unitarity of $\wb$ leads to a constraint on the
generalized almost-complex structure $\wm$, namely,
 \eq \label{qeq3} \widetilde{M}^\dag\equiv\widetilde{M}^{\ast T}=
  \widetilde{M}^{-1}=-\widetilde{M} \Rightarrow q_{ij}^\ast
  q_{ij}=1,
\en where the symbol $\ast$ denotes the complex conjugation and the
symbol $T$ denotes the transpose operation.

As $J$ is a half-integer, we obtain solutions for equations
(\ref{qeq1}) and $(\ref{qeq3})$ in terms of independent $(J+\frac 1
2)$ number of angle parameters $\varphi_{J}$, $\varphi_{J-1}$,
$\cdots$, $\varphi_{\frac 1 2}$,
  \eq q_{lm}=e^{i\frac {\varphi_l +\varphi_m} 2}, \,\,
 \varphi_{-l}=-\varphi_l, \qquad  0\le l\le J, \en
where the method of separation of variables has been used since one
can choose $q_l=e^{i\varphi_l}$ and then $q_{lm}=q_{l} q_m$.

 For example, deformation parameters in the unitary generalized Bell matrix
 $\widetilde{B}^{\frac 1 2\frac 1 2}$ are calculated to be
  \eqa
  q_{\frac 1 2\frac 1 2}=e^{i\varphi},\,\, q_{-\frac 1 2-\frac 1 2}
  =e^{-i\varphi},\,\,q_{\frac 1 2-\frac 1 2}=q_{-\frac 1 2\frac 1 2}=1,
  \ena
which are the same as those presented \cite{yong1,yong2,yong3}, and
deformation parameters of the generalized Bell matrix
$\widetilde{B}^{\frac 3 2\frac 3 2}$ have the form,
 \eqa
 &&  q_{\frac 3 2 \frac 3 2}=e^{i\varphi_1},\,\,
  q_{\frac 3 2 \frac 1 2}=e^{i\frac {\varphi_1+\varphi_2} 2},\,\,
  q_{\frac 3 2 -\frac 1 2}=e^{i\frac {\varphi_1-\varphi_2} 2},\,\,
  q_{\frac 3 2 -\frac 3 2}=1,
   \nonumber\\
 & &  q_{\frac 1 2 \frac 3 2}=e^{i\frac {\varphi_1+\varphi_2} 2},\,\,
 q_{\frac 1 2 \frac 1 2}=e^{i\varphi_2},\,\, q_{\frac 1 2 -\frac 1 2}=1,\,\,
 q_{\frac 1 2 -\frac 3 2 }=e^{i\frac {\varphi_2-\varphi_1} 2}.
 \ena

In the $2^{2n}$-dimensional\footnote{Here we have $2^{2n}=(2J+1)^2$,
for example, $n=1, J=\frac 1 2$ and $n=2, J=\frac 3 2$, see
(\ref{nj}). } vector space, the generalized almost-complex structure
$\widetilde{M}$ is found in this paper to satisfy algebraic
relations, \eqa \label{mal} & & \widetilde{M}^2=-1\!\! 1,\qquad
\widetilde{M}_{i\pm 1} \widetilde{M}_i =-\widetilde{M}_i
\widetilde{M}_{i\pm 1},\nonumber\\
&& \widetilde{M}_{i }\widetilde{M}_j =\widetilde{M}_j
\widetilde{M}_{i},\qquad  |i-j|\ge 2,\, i,j\in {\mathbb N}, \ena
which defines an algebra obviously different from the
Temperley--Lieb algebra \cite{lieb} or the symmetric group algebra
and where deformation parameters $q_{ij}$ satisfy
 \eq
q_{ij} q_{-i-j}=1, \qquad q_{ij} q_{-ij}=q_{jk} q_{j-k}.
 \en
 With the help of this algebra (\ref{mal}), the generalized Bell matrix
$\wb$ can be easily proved to satisfy the braided YBE (\ref{bgr}) in
the way
 \eq
 \widetilde{B}_i \widetilde{B}_{i+1} \widetilde{B}_i =2
\widetilde{M}_i+2 \widetilde{M}_{i+1}+\widetilde{M}_i
\widetilde{M}_{i+1} +\widetilde{M}_{i+1}\widetilde{M}_i
 =\widetilde{B}_{i+1} \widetilde{B}_{i} \widetilde{B}_{i+1}.
 \en

Additionally, the generalized almost-complex structure $\wm$ and the
permutation operator $P$ satisfy the following algebraic relation
 \eq P_i \wm_{i+1} P_i =P_{i+1} \wm_i P_i, \qquad
    P=\sum_{ij}|ij\rangle\langle j i|, \en which is underlying
algebraic relations of the virtual braid group, i.e., the braid
$\wb$ and permutation $P$ forming a unitary virtual braid
representation, see \cite{yong4,yong5}.

\subsection{New type of solution of QYBE via parameterization}

Similar to the formalism of the rational solution of the QYBE
(\ref{qybe}), \eq \check{R}_{rational}(u)=1\!\! 1 +u P, \qquad
P^2=1\!\! 1 \en where $P$ is a permutation matrix, we obtain a
solution of the QYBE in terms of the generalized almost-complex
structure, \eq
 \widetilde{\check{R}}(u)=1\!\! 1 + u \wm
\en satisfying the following equation of Yang--Baxter type,
  \eq
  \label{mybe}
 \check{R}_i(u)\,\check{R}_{i+1}(\frac {u+v} {1+u v})\,
 \check{R}_{i}(v)= \check{R}_{i+1}(v)\,\check{R}_{i}( \frac
{u+v} {1+u v})\,\check{R}_{i+1}(u),
 \en
which has been exploited \cite{yong2} and where new spectral
parameters $u, v$ are related to original spectral parameters $x,y$
in the way
 \eq u=\frac {1-x} {1+x},\qquad v=\frac
{1-y} {1+y},\qquad \frac {1-x y} {1+x y}=\frac {u+v} {1+u v}.\en

Via a further parametrization of spectral parameters $u, v$ in terms
of angle variables $\Theta_1, \Theta_2$, \eq
u=-\sqrt{-1}\tan\Theta_1,\,\, v=-\sqrt{-1}\tan\Theta_2,\, \,\frac
{u+v} {1+u v}=-\sqrt{-1}\tan(\Theta_1+\Theta_2),
 \en
the modified Yang--Baxter equation (\ref{mybe}) has the ordinary
form
 \eq
\check{R}_i(\Theta_1)\,\check{R}_{i+1}(\Theta_1+\Theta_2)\,
 \check{R}_{i}(\Theta_2)= \check{R}_{i+1}(\Theta_2)\,
 \check{R}_{i}(\Theta_1+\Theta_2)\,\check{R}_{i+1}(\Theta_1).
 \en
with the solution given by \eq \widetilde{\check{R}}(\Theta)=1\!\!
1-\sqrt{-1}\tan\Theta \wm, \qquad \textrm{or}\qquad
\widetilde{\check{R}}(\Theta^\prime)=1\!\! 1+\tanh\Theta^\prime
\wm.\en Note that physical models underlying this type of solution
of QYBE in terms of the almost-complex structure will be discussed
and submitted elsewhere.

\section{Projectors, diagonalization and geometry}

This section and the next one are aimed at introducing several
selective topics directly using the generalized Bell matrix and the
generalized almost-complex structure, for example, associated
noncommutative geometry, quantum algebra and FRT dual algebra.

\subsection{Projectors and  Yang--Baxterization}

In terms of $\wm$, two projectors $\widetilde{P}_+$ and
$\widetilde{P}_-$ are defined by \eq \label{proj} \pw_+=\frac 1 2
(1+\sqrt{-1} \wm), \qquad \pw_-=\frac 1 2 (1-\sqrt{-1} \wm) \en
satisfying basic properties of two mutually orthogonal projectors,
 \eq
\pw_+ + \pw_-=1\!\! 1,\,\,\pw^2_\pm=\pw_\pm,\,\,\pw_+\pw_-=0.
 \en

The generalized Bell matrix $\wb$ has two distinct eigenvalues
$e^{\pm i \frac \pi 4}$ and it satisfies the same characteristic
equation as (\ref{char}), \eq (\widetilde{B}-\lambda_- 1\!\! 1)
(\widetilde{B}- \lambda_+ 1 \!\! 1)=0, \qquad \lambda_+=e^{-i\frac
\pi 4},\,\,\, \lambda_-=e^{i\frac \pi 4}.
 \en
With the projectors $\pw_\pm$ and eigenvalues $\lambda_\pm$, the
generalized Bell matrix and its inverse have the forms
 \eq
\wb=\lambda_+ \pw_+ + \lambda_- \pw_-, \qquad \wb^{-1}=\lambda_-
\pw_+ + \lambda_+ \pw_-.
 \en

Using Yang--Baxterization \cite{yong2}, the $\check{R}(x)$-matrix
 as a solution of the QYBE (\ref{qybe}) has a form similar to (\ref{rx}),
 \eq
 \widetilde{\check{R}}(x) =(\lambda_+ +\lambda_- x)\pw_+ +
  (\lambda_- +\lambda_+ x)\pw_- =\wb +x \wb^{-1}, \en
and hence the corresponding Schrodinger equation also has a similar
form to (\ref{scr1}) or (\ref{scr2})  except that the Hamiltonian is
determined by $\wm$ instead of $M$.

\subsection{Diagonalization of the generalized Bell matrix }

The diagonalization of the generalized Bell matrix $\wb$ can be
performed by a unitary matrix $D$ via the following unitary
transformation, \eq
 D\wb D^\dag =\frac 1 {\sqrt 2} Diag(1+\sqrt{-1}, \cdots, 1-\sqrt{-1})
 \en
where the diagonal matrix $Diag$ has the same number of matrix
entries $1+\sqrt{-1}$ as $1-\sqrt{-1}$. Assume $D$ to have the form
by a Hermitian unitary matrix $N$,
 \eq
 D=\frac 1 {\sqrt 2} (1\!\! 1 + \sqrt{-1} N), \qquad N^\dag=N, \qquad
 N^2=1\!\! 1
 \en
 and then this matrix $N$ is found to satisfy an additional condition,
 \eq
 N\wm=-\wm N =Diag(1,-1, \cdots, 1, -1),
 \en
 where the diagonal matrix $Diag$ has the same number of matrix entries $1$
 as $-1$ but the ordering between $1$ and $-1$ is not fixed.

After some algebra, one type of formalism of the matrix $N$ is given
by
  \eq
N_{ij}^{kl} =f(i)q_{ij} \delta_i^{-k} \delta_j^{-l}, \qquad
f(i)f(i)=1,\,\, f(i)=f(-i)=f^\ast(i)
  \en
where $q_{ij}$ are the same as unitary deformation parameters
$q_{ij}$ in the generalized Bell matrix $\wb$. This matrix $N$
brings about the diagonalization form of $\wb$, \eq
 (D\wb D^\dag)_{ij}^{mn} =\frac 1 {\sqrt 2} (1+\sqrt{-1}
 f(i)\epsilon(-i)) \delta_i^m\delta_j^n.
\en in which setting $f(i)=\epsilon(-i)$, $i>0$ and
$f(i)=\epsilon(i)$, $i<0$ leads to
 \eq
 D\wb D^\dag =\frac 1 {\sqrt 2} Diag(\underbrace{1+\sqrt{-1},
  \cdots,1+\sqrt{-1}}_{2^{N-1}},
 \underbrace{1-\sqrt{-1},\cdots, 1-\sqrt{-1}}_{2^{N-1}}).
 \en

In the four dimensional case, for example, the Bell matrix $B_4$ is
diagonalized in the way,
 \eq
 D_{4}B_4 D^\dag_{4}= \frac 1 {\sqrt 2}
  Diag(1-\sqrt{-1},1+\sqrt{-1},1-\sqrt{-1},1+\sqrt{-1}),\,\,
    N_4=-\sigma_y\otimes \sigma_y,
 \en
 where a note is added that $B_4$ can be also diagonalized by unitary
 transformations of the Malkline matrix (or the magic matrix)
 {\cite{fujii1, fujii2, makhlin} or the diagonaliser \cite{acdm1}, and the following
 generalized Bell matrix $\wb_4$ can be diagonalized with a given $N_{4,1}$,
  \eqa
 & & \wb_4=\frac 1 {\sqrt 2}\left( \begin{array}{cccc}
  1 & 0 &  0 & q \\ 0 & 1 & 1 & 0 \\ 0 & -1 & 1 & 0 \\ -q^{-1} & 0 & 0 & 1
  \end{array} \right),\,\,
  N_{4,1}= \left( \begin{array}{cccc}
   0 & 0  &  0 & - q \\ 0 & 0 & -1 & 0 \\
   0 & 1 & 0 & 0 \\  q^{-1} & 0 & 0 & 0
    \end{array} \right), \nonumber\\
  & & D_{4,1} \wb_4 D_{4,1}=\frac 1 {\sqrt 2}
   Diag(1+\sqrt{-1},1+\sqrt{-1},1-\sqrt{-1},1-\sqrt{-1}).
 \ena

As a remark, calculation for noncommutative geometry and quantum
algebra associated with the generalized Bell matrix can be greatly
simplified once the above diagonalization procedure is exploited.

\subsection{Associated noncommutative geometry}

With the help of the standard procedure of setting up associated
noncommutative geometry with a given braid representation
\cite{wz,madore}, we denote coordinate operators $X$ and
differential operators $\xi$ in the way \eq
 X^T=(x_1,x_2\cdots, x_{2^{N}}), \qquad \xi^T=(\xi_1,\xi_2,\cdots, \xi_{2^{N}})
\en and demand them to satisfy constraint equations,
 \eq
\pw_-(X\otimes X)=0, \qquad \pw_+ (\xi\otimes \xi)=0, \qquad
X\otimes\xi=(\mu \pw_+-1\!\! 1)(\xi\otimes X)
 \en
where $\mu$ is a free parameter. These three equations can be chosen
in the second way by exchanging $\pw_+$ with $\pw_-$ but this
approach is omitted here for simplicity.

Hence noncommutative differential geometry generated by $X$ and
$\xi$ is  essentially determined by the following equations in terms
of the generalized almost-complex structure $\wm$,
 \eqa
&& X\otimes X=\sqrt{-1}\wm (X\otimes X), \qquad
\xi\otimes\xi=-\sqrt{-1}\wm(\xi\otimes\xi), \nonumber\\
&& X\otimes\xi=(\frac \mu 2-1)\xi\otimes X +\frac \mu 2 \sqrt{-1}
\wm (\xi\otimes X),
 \ena
which have the following formalisms of component,
 \eqa
 && x_i x_j =\sqrt{-1}\epsilon(i) q_{ij} x_{-i} x_{-j}, \qquad
   \xi_i \xi_j =-\sqrt{-1}\epsilon(i) q_{ij} \xi_{-i} \xi_{-j},
   \nonumber\\
 && x_i \xi_j =(\frac \mu 2-1) \xi_i x_j +
  \frac \mu 2 \sqrt{-1} \epsilon(i) q_{ij} \xi_{-i} x_{-j}.
 \ena
with the significant geometry at $\mu=2$. Note that noncommutative
plane related to the Bell matrix $B_4$ has been briefly discussed
\cite{acdm2}.

\section{Quantum algebra via the FRT procedure}

For a given solution $\check{R}$ of the braided YBE (\ref{bgr}),
there is a standard procedure \cite{rtf,faddeev} using the
$\check{R}TT$ relation and $\check{R}LL$ relations to respectively
define associated quantum algebra and FRT dual algebra. In this
section, we sketch quantum algebra and FRT dual algebra specified by
the $\wb TT$ relation and $\wb LL$ relations.

\subsection{Quantum algebra using the $\wm TT$ relation}

In the well known $\check{R}TT$ relation: $\check{R}(T\otimes
T)=(T\otimes T)\check{R}$, matrix entries of the $T$-matrix are
assumed to be non-commutative operators. Here the $\check{R}$-matrix
is the generalized Bell matrix $\wb$, and the $\wb TT$ relation is
essentially determined by $\wm TT$ relation,
 \eq
 \wb(T\otimes T)=(T\otimes T)\wb \Rightarrow
  \wm(T\otimes T)=(T\otimes T)\wm,
 \en
where $\wm$ is a $2^{2n}\times 2^{2n}$ matrix and $T$ is a
$2^{2n-1}\times 2^{2n-1}$ matrix. In terms of matrix entries of
$\wm$, $T$ and the convention $(A\otimes B)_{ij,kl}\equiv A_{ik}
B_{jl}$, the $\wm TT$ relation has the following component
formalism,
 \eqa
 \label{mtt}
 & & T_{i_1-i_{2}} T_{j_1 -j_2}+\epsilon(i_1)\epsilon(i_2) q_{i_1j_1} q_{i_2j_2}
 T_{-i_1i_2} T_{-j_1j_2}=0, \nonumber\\
 & & i_1, i_2, j_1, j_2=J, J-1, \cdots,-J.
 \ena

Note that this $\wm TT$ relation (\ref{mtt}) has eight simplified
equations of component,
 \eqa
 && T_{i i} T_{i i}=T_{-i-i} T_{-i-i},\qquad\qquad
  T_{i i}T_{-i-i}=T_{-i-i} T_{i i},
 \nonumber\\
 && T_{i-i} T_{i-i}=-q^2_{ii} T_{-ii} T_{-ii},
 \qquad T_{i-i} T_{-ii} =-T_{-ii} T_{i-i},
 \nonumber\\
 && T_{i i} T_{i -i}=q_{ii} T_{-i-i} T_{-ii},
 \qquad \,\,\, T_{i i} T_{-ii}=q_{-i-i}T_{-i-i}
 T_{i-i}, \nonumber\\
 && T_{i-i} T_{i i}=-q_{ii} T_{-ii} T_{-i-i},
 \qquad T_{i-i} T_{-i -i}=-q_{ii} T_{-ii} T_{i i}
 \ena
which determine the quantum algebra related to $\wb_4$.

With the help of a new $\wt$-matrix given by
 \eq
 \wt_{ij}=\epsilon(i)T_{ij} + T_{-i-j}, \qquad
 \wt_{-i-j}=-\epsilon(i)T_{-i-j} + T_{ij},
 \en
where $q_{ij}$ is chosen to be unit for convenience, the $\wm TT$
relation (\ref{mtt}) is replaced by the $\wm \wt \wt$ relation
having the algebraic relations, \eqa &&
\wt_{i_1-i_2}\wt_{j_1-j_2}=-\wt_{-i_1i_2}\wt_{-j_1j_2},
\,\,\,\,\,\,\, \textrm{if}\qquad \epsilon(i_1)\epsilon(i_2)=1,\,\,\,
\epsilon(i_2)\epsilon(j_1)=1,
\nonumber\\
&& \wt_{i_1-i_2}\wt_{-j_1j_2}=\wt_{-i_1i_2}\wt_{j_1-j_2},
\qquad\,\textrm{if}\qquad \epsilon(i_1)\epsilon(i_2)=1,\,\,\,
\epsilon(i_2)\epsilon(j_1)=-1, \nonumber\\
& & \wt_{i_1-i_2}\wt_{j_1-j_2}=\wt_{-i_1i_2}\wt_{-j_1j_2},
\qquad\textrm{if}\qquad \epsilon(i_1)\epsilon(i_2)=-1,\,\,
 \epsilon(i_2)\epsilon(j_1)=1, \nonumber\\
&& \wt_{i_1-i_2}\wt_{-j_1j_2}=-\wt_{-i_1i_2}\wt_{j_1-j_2},
\,\,\,\,\,\textrm{if}\,\,\,\,  \epsilon(i_1)\epsilon(i_2)=-1, \,\,
\epsilon(i_2)\epsilon(j_1)=-1,
 \ena
which leads to four simplest algebraic relations, \eqa
\wt_{i-i}^2=0,\,\,\wt_{ii}\wt_{-i-i}=0,\,\,
 \wt_{ii}\wt_{-ii}=0,\,\, \wt_{i-i}\wt_{ii}=0.
\ena

\subsection{Example: the quantum algebra from the $\wb_4TT$ relation}

The $\wb_4$-matrix and $T$-matrix take the forms,  \eq
\wb_4=\left(\begin{array}{cccc}
1 & 0 & 0 & q \\
0 & 1 & 1 & 0 \\
0 & -1 & 1 & 0 \\
-\, q^{-1} & 0 & 0  & 1
\end{array}\right), \qquad T=\left(\begin{array}{cc}
   \hat{a}  &  \hb  \\
   \hc  &  \hd   \\
 \end{array}\right),  \en
and the $\wb_4 TT$ relation leads to the quantum algebra generated
by $\hat{a}, \hb, \hc, \hd$ satisfying algebraic relations, \eqa
\label{algebra}
 \ha \ha &=&  \hd \hd, \,\, \ha \hb = q \hd \hc,\,\,\hb \hb =  - q^2 \hc
 \hc,\,\,\ha \hc =q^{-1} \hd \hb, \nonumber\\
 \ha \hd &=& \hd \ha,\,\, \hb \ha = - q \hc \hd, \,\,
 \hb \hc = - \hc \hb, \,\,\hc \ha = - q^{-1} \hb \hd, \ena
where the deformation parameter $q$ can be absorbed into the
generator $\hc$ by a rescaling transformation. With the new
operators $\widetilde{\ha}, \widetilde{\hb}, \widetilde{\hc},
\widetilde{\hd}$ \cite{acdm3} specified by \eq
\widetilde{\ha}=\ha+\hd,\,\, \widetilde{\hb}=\hb+\hc,\,\,
\widetilde{\hc}=\hb-\hc,\,\, \widetilde{\hc}=\ha-\hd,
  \en
the above algebraic relations have a very simplified formalism, \eq
 \widetilde{\ha} \widetilde{\hd} =\widetilde{\hd}\widetilde{\ha}=0, \,\,
 \widetilde{\hb}\widetilde{\hb} =\widetilde{\hc}\widetilde{\hc}=0,\,\,
 \widetilde{\ha}\widetilde{\hc} =\widetilde{\hd}\widetilde{\hb}=0,\,\,
 \widetilde{\hb}\widetilde{\ha} =\widetilde{\hc}\widetilde{\hd}=0.
\en

Note that the quantum algebra from the $B_4TT$ relation and its
representation theory has been presented \cite{acdm3}, while the
same quantum algebra from the $\wb_4 TT$ relation, interesting
algebraic structures underlying its representation and its natural
connection to quantum information theory has been explored
\cite{yong3}.

As a remark, quantum algebra obtained from $\wb TT$ relation may be
higher-dimensional representations of that algebra given by
$\wb_{4}TT$ relation, see \cite{yong8}.

\subsection{FRT dual algebra using the $\wm LL$ relations}

The $\check{R}LL$ relations determining the FRT dual algebra can be
derived from the generalized $\check{R}TT$ relation which relies on
the spectral parameter,
 \eq
 \check{R}(xy^{-1})(L(x)\otimes L(y))
  =(L(y)\otimes L(x))\check{R}(xy^{-1}).
 \en
Assume the $L(x)$-matrix to have a similar form to $\wb(x)$, \eq
L(x)= L^+ + x\, L^{-},\qquad \wb(x)= \wb + x \wb^{-1}, \en and this
leads to the $\wb LL$ relations for the FRT dual algebra,
 \eq
 \wb (L^\pm\otimes L^\pm) = (L^\pm\otimes L^\pm) \wb,\qquad
 \wb (L^+ \otimes L^-) = (L^- \otimes L^+) \wb,
   \en
where matrix entries of $L^\pm$ are non-commutative operators. These
$\wb LL$ relations are found to be essentially $\wm LL$ relations,
\eqa & & \wm (L^\pm\otimes L^\pm) = (L^\pm\otimes
L^\pm)\wm,\nonumber\\
& &  L^+ \otimes L^- - L^- \otimes L^+ + \wm (L^+ \otimes L^-)-(L^-
 \otimes L^+) \wm=0,
\ena which have the component formalisms,  \eqa
 && L^\pm_{i_1-i_{2}} L^\pm_{j_1 -j_2}+
  \epsilon(i_1)\epsilon(i_2) q_{i_1j_1} q_{i_2j_2} L^\pm_{-i_1i_2}
   L^\pm_{-j_1j_2}=0,\nonumber\\
 && L^+_{i_1i_2} L^-_{j_1j_2} - L^-_{i_1i_2}L^+_{j_1j_2} +\epsilon(i_1)
q_{i_1j_1} L^+_{-i_1i_2}L^-_{-j_1j_2} +
\epsilon(i_2)q_{-i_2-j_2} L^-_{i_1-i_2}L^+_{j_1-j_2}=0, \nonumber\\
 &&     i_1, i_2, j_1, j_2=J, J-1, \cdots,-J,
 \ena

Note that the FRT dual algebra for the Bell matrix $B_4$ has been
given \cite{acdm1} and a quotient algebra of this FRT dual algebra
with the quotient condition $L^+ \otimes L^- = L^- \otimes L^+$ has
been presented \cite{yong3}. Also, in view of \cite{acdm1,acdm3,
yong3}, further research is needed to construct representation
theories and seek interesting algebraic structures underlying them
for these quantum algebra and FRT dual algebra.

\section{Concluding remarks and outlooks}

This paper is motivated by the recent study
\cite{dye,kauffman1,yong1,yong2,yong3}, and it sheds a light on
further research for unraveling  deep connections among quantum
information theory, Yang--Baxter equation and complex geometry. We
find that the GHZ states can be yielded by the Bell matrix on the
product base and prove that the generalized Bell matrix of the type
$2^{2n}\times 2^{2n}$ forms a unitary braid representation with the
help of the algebra generated by the almost-complex structure. The
algebraic and diagrammatic proofs for the generalized Bell matrix of
the type $2^{2n+1}\times 2^{2n+1}$ satisfying the braided YBE
together with other interesting result will be submitted
\cite{yong8}.

Besides what we have done in the present paper, there still remain
many meaningful topics worthwhile to be explored.  For example,
almost-complex structure, classical YBE and symplectic geometry;
construction of a universal $R$-matrix \cite{kassel} in terms of the
generators of the algebra from the $\wb_4 TT$ relation; Yangian,
Yang--Baxter equation and quantum information; new quantum algebra
obtained by exploiting methodologies for the Sklyanin algebra
\cite{sklyanin1,sklyanin2} to the generalized Bell matrix. The most
important thing (at least for the authors) is still to look for
further connections among physics, quantum information and the YBE.

\section*{Acknowledgments}
Both authors are grateful to L.H. Kauffman and N. Jing for fruitful
collaborations, and they thank C.M. Bai and J.L. Chen for
stimulating discussions. The first author thanks K. Fujii for
helpful references and A. Chakrabarti for email correspondence on
the $RLL$ relations for the FRT dual algebra, and especially thanks
Y.S. Wu for crucial and stimulating comments on this manuscript.
This work is in part supported by NSFC grants (-10605035) and SRF
for ROCS, SEM.

\end{document}